\documentstyle[sprocl,epsf]{article}
\def\be{\begin{equation}}
\def\ee{\end{equation}}
\def\ba{\begin{eqnarray}}
\def\ea{\end{eqnarray}}
\def\newblock{}

\begin{document}
\vspace*{-2cm}
\begin{flushright}
CERN-TH/96-280\\
hep-ph/9610247\\
September 1996\\
\end{flushright}

\title{FINITE TEMPERATURE EFFECTIVE THEORIES}

\author{M. E. SHAPOSHNIKOV}

\address{Theory Division, CERN, CH-1211 Geneva 23, Switzerland}


\maketitle\abstracts{Lecture Notes, Summer School on Effective 
Theories and Fundamental Interactions, Erice, 1996. I describe 
the construction of effective field theories for equilibrium 
high-temperature plasma of elementary particles.}

\section{Motivation}
High-temperature and dense matter of elementary particles appears in
several areas of physics. The first and most familiar example is the
Universe at the early stages of its expansion. The Big Bang theory
(for reviews see, e.g. \cite{zeldbook,koltur:rev}) states that the
Universe was hot and dense in the past, with a temperature ranging
from a few eV up to the Planck scale $\sim M_{Pl} \sim 10 ^{19}$ GeV.
Another place, where matter exists in extreme conditions of high
densities, can be created in the laboratory. Namely, in heavy-ion
collisions extended dense fireballs of nuclear matter are created,
with the energy density exceeding the QCD scale $\sim 10\cdot (200 \mbox
MeV)^4$.

{}From a practical point of view, there is quite a wide area of
applications of finite-temperature field theory to cosmology and
laboratory experiments. The high-temperature phase transitions,
typical for grand unified theories (GUTs), may be important for
cosmological inflation and primordial density fluctuations. 
Topological defects (such as monopoles, strings, domain walls) can
naturally arise at the phase transitions and influence the
properties of the Universe we observe today. The first-order
electroweak (EW) phase transition is a crucial element for
electroweak baryogenesis; it may also play a role in the formation of
the magnetic fields observed in the Universe. The QCD phase
transition and properties of the quark-gluon plasma are essential for
the understanding of the physics of heavy-ion collisions. The QCD phase
transition in cosmology may influence the spectrum of the density
fluctuations relevant to structure formation.

In most cosmological applications, the deviations from thermal
equilibrium play a most important role. For example, the
concentration of primordial monopoles depends a lot on the dynamics
of the grand unified phase transition; the baryonic asymmetry,
produced in GUTs or in the EW theory, depends on the rate of the Universe
expansion or other time-dependent phenomena. The signatures of
the heavy-ion collisions are greatly influenced by the
non-equilibrium dynamics.  The study of non-equilibrium properties of
the plasma is very hard. To understand the thermal non-equilibrium
better, the systems in thermal equilibrium should be completely
understood. Without this understanding, we cannot even say what a
deviation from thermal equilibrium is. After the structure of the
ground state is found, small deviations from thermal equilibrium
can be treated by this or that perturbative method.

The topic of my lecture here is the study of the {\em equilibrium}
properties of the plasma at high temperatures. As we will see, even
this problem in gauge theories is highly non-trivial and requires the
use of effective field theory methods.

The study of dense matter is interesting in itself from a theoretical
point of view. At very small temperatures and densities the plasma
can be considered as a collection of weakly interacting individual
particles; its thermodynamical properties are close to those of the
ideal Bose or Fermi gas. When the temperature or density increases,
this is not true any longer, and the collective properties of plasma
become important. The change of temperature and/or density of the
system may induce phase transformations in the system. For example,
the Higgs phase of the electroweak theory, realized at small
temperatures, may be transformed into a ``symmetric" phase
\cite{kir,kirlin,doljack,weitemp} (it is often said that the
symmetries are ``restored" at high temperatures). Another example is
the theory of strong interactions. At small temperatures the
confinement phase of QCD is realized, while at higher temperatures
the confinement is believed to be absent \cite{pol,sussk}, so that a
better description of the system may be achieved in terms of quarks
and gluons. Thus, the study of finite-temperature field theory can be
considered as a theoretical laboratory to test our understanding of
the Higgs mechanism in the EW theory, confinement and chiral symmetry
breaking in QCD, etc.

The plan of the lecture is as follows. In section 2 we discuss the
accuracy of the equilibrium approximation in cosmology. In section 3
we introduce the main definitions of finite-temperature field theory,
and in section 4 we explain why ordinary perturbation theory breaks
down at high temperatures. In section 5 we explain the main idea of
an effective field theory approach and in section 6 apply it to the
different physical theories. In section 7 we discuss phase transition
in the EW theory. Section 8 is the conclusion. Our discussion is
carried out mainly on the qualitative level, many technical details
can be found in the original papers cited below.

\section{Equilibrium approximation}
In loose terms, the thermal equilibrium approximation may be valid if
the system is considered after some time substantially exceeding the
typical equilibration time. Let us see how this general rule works
for the expanding Universe.

The measure of deviation from the thermal equilibrium is the ratio of
two time scales. The first one is the rate of the Universe expansion,
given by the inverse age of the Universe $t_U^{-1}$: $ t_U=
\frac{M_0}{T^2}$. Here $M_0 = M_{Pl}/1.66 N^{\frac{1}{2}}\sim
10^{18}$ GeV, and $N$ is the effective number of the massless degrees
of freedom. The expansion rate of the Universe is a unique
non-equilibrium parameter of the system (at least in the absence of
different phase transitions). The second time scale (different for
different types of interaction) is a typical reaction time, given by 
$(\tau_{reaction})^{-1} \sim \langle \sigma n v \rangle, $ where
$\sigma$ is the corresponding cross-section, $n$ is the particle
concentration and $v$ is the relative velocity of the colliding
particles.

As an example, let us consider the Universe at the electroweak epoch,
at temperatures $T\sim m_W$. The fastest reactions are those
associated with strong interactions (e.g. $q\bar{q} \rightarrow GG$);
their rate is of the order of $(\tau_{strong})^{-1} \sim \alpha_s^2
T$. The typical weak reactions, say $e\nu \rightarrow e\nu$, occur at
the rate $(\tau_{weak})^{-1} \sim \alpha_W^2 T$, and the slowest
reactions are those involving chirality flips for the lightest
fermions, e.g. $e_R H\rightarrow \nu W$ with the rate
$(\tau_{e})^{-1} \sim f_e^2 \alpha_W T$, where $f_e$ is the electron
Yukawa coupling constant. Now, the ratio $\frac{\tau_i}{t_U}$ varies
from $10^{-14}$ for the fastest reactions to $10^{-2}$ for the
slowest ones. This means that particle distribution functions of
quarks and gluons, intermediate vector bosons, Higgs particle and
left-handed charged leptons and neutrino are equal to the equilibrium
ones with an accuracy better than $10^{-13}$; the largest deviation
from thermal equilibrium ($\sim 10^{-2}$) is being expected for the
right-handed electron.

These estimates show that the equilibrium description of the Universe
is a very good approximation at the electroweak scale. At other
temperatures the situation may not be so optimistic. For instance, if
the Universe was as hot as, say, $10^{17}$ GeV, then the equilibrium
description of the (grand unified) interactions would be
questionable, since the ratio $\frac{\tau_i}{t_U}$ would be of the
order of $1$. By equating the rate of the Universe expansion with the
rate of this or that reaction, one can estimate a range of
temperatures where the process under consideration was in thermal
equilibrium. For example, the rate of the electromagnetic
interactions exceeded the rate of the Universe expansion at (few eV) 
$< T < \alpha^2 M_{Pl} \sim 10^{15}$ GeV, weak interactions were in
thermal equilibrium at temperatures up to the nucleosynthesis
temperature $\sim$ (few keV), etc.

The same type of consideration can be carried out for the heavy-ion
collisions. Again, if the thermal equilibration time is much smaller
than the time the fireball exists, equilibrium thermodynamics can be
applied to the description of the extreme state of matter appearing
as an intermediate stage of the collision process. Since strong dynamics
is involved, the estimates of the corresponding time scales are much
less certain, but quite encouraging \cite{Eskola:1988,ion1,ion2} (for
a review see, e.g. \cite{satz}).

The general conclusion is that the equilibrium approximation is valid
in a wide range of physical situations in cosmology and, less
certainly, in the laboratory.

\section{The basics of finite-temperature field theory}
This section is a short introduction to finite-temperature field
theory. More details can be found in a number of excellent reviews
and books, see, e.g. refs. \cite{Kir76,Li90,kapusta}.

If some system is described by the Hamiltonian $H$, and there are
several conserved charges $Q_i$, $[Q_i,H]=0$, then the thermal
equilibrium state of the system is described by the density matrix
$\rho$,
\be
\rho= \frac{1}{Z} \exp\left(-\frac{1}{T}(H + \sum_i \mu_i Q_i)\right),
\label{dens}
\ee
where $\mu_i$ is a set of chemical potentials. The parameter $Z$ is
nothing but the statistical sum of the system:
\be
Z = Tr \left[\exp\left(-\frac{1}{T}(H + \sum_i \mu_i Q_i)\right)\right].
\ee
In what follows we constrain the general situation to the case when
all chemical potentials of the system are equal to zero. In this
particular case the statistical sum is related to the density of the
free energy $F$ of the system  through
\be
Z= \exp\left(-\frac{F V}{T}\right),
\ee
where $V$ is the volume of the system. The analogy of expression
(\ref{dens}) with the quantum-mechanical time evolution operator $\exp
(-i H t)$ allows us to write down a functional-integral representation
for the statistical sum:
\be
Z = \int D\phi D\Psi \exp(-S_E),
\ee
where the integral is taken over all bosonic ($\phi$) and fermionic
($\Psi$)  fields, and $S_E$ is Euclidean action for the system,
defined on a finite ``time" interval $0< \tau < \beta =\frac{1}{T}$,
\be
S_E = \int_0^\beta d \tau \int d^3 x L,
\ee
where $L$ is the Euclidean Lagrangian density. The bosonic fields,
entering the functional integral, obey periodic boundary conditions
with respect to imaginary time, $\phi(0,x) = \phi(\beta, x)$, and
fermionic fields are antiperiodic, $\Psi(0,x) = -\Psi(\beta, x)$.
The case of gauge theories requires the ordinary gauge fixing and
introduction of ghost fields. In spite of the anticommuting character
of the ghost fields, they obey periodic boundary conditions.

The formal analogy with zero-temperature field theory allows the
introduction of an important notion of the so-called imaginary time (as
opposed to real time), or Matsubara Green's functions. For example,
the bosonic Green functions are defined as
\be
G(\tau_1,x_1,...,\tau_n,x_n)= \frac{1}{Z}
\int \phi(\tau_1,x_1)...\phi(\tau_n,x_n) D\phi D\Psi \exp(-S_E).
\ee
Fermionic Green's functions are derived by a simple replacement of
the bosonic fields by fermionic fields.

The construction described above is the basis of the statement that the
finite temperature {\em equilibrium} field theory is equivalent to
the Euclidean field theory defined on a finite ``time" interval. Thus,
many methods developed for the description of zero-temperature quantum
field theory (e.g. perturbation theory, semi-classical analysis,
lattice numerical simulations) can be easily generalized to the non-zero
temperature case.

For example, perturbation theory at finite temperatures looks
precisely like perturbation theory at $T=0$ with substitutions of
quantities associated with the zero component of 4-momentum $p$, as
follows:
\[
p_0 \rightarrow i\omega,
\]
\be
\int dp_0 \rightarrow 2 \pi i T \sum_{\omega},
\ee
\[
\delta(p_0) \rightarrow (2 \pi i T)^{-1} \delta_{\omega,0},
\]
where the discrete variable $\omega$ is $2 \pi n T$ for the bosons
and $(2n +1) \pi T$ for fermions, with $n$ being an integer number.
The finiteness of the time interval makes the energy variable 
discrete, since the Fourier integral used at zero temperature is
substituted by the Fourier sum.

The equilibrium properties of a plasma are completely defined by the
statistical sum and by the set of Green's functions. Thus the problem
of equilibrium statistics is to compute these quantities reliably.

\section{The breakdown of perturbation theory}
In this lecture we will constrain ourselves to the theories of the
following  type. We will require that the running coupling constants
of the theory, taken at the scale of the order of temperature, are
small. This class is sufficiently wide and includes many interesting
cases. The simplest example is a scalar theory
\be
L= \frac{1}{2}(\partial_\mu \phi)^2 + \frac{1}{2}m^2 \phi^2 +
\frac{\lambda}{4} \phi^4
\label{phi4}
\ee
with $\lambda \ll 1$. Since the scalar self-coupling is not
asymptotically free, the temperature of the system is assumed to be
much smaller than the position of the Landau pole. Another example is
the electroweak theory, and another is QCD at sufficiently large
temperatures ($T \gg 100$ MeV). Many GUTs also fall
in this class.

An aim of this section is to show that the straightforward or
modified perturbation theory fails in describing certain details of
the properties of high-temperature plasma. We will also discuss here
difficulties in putting the whole problem on the lattice. We will use
here fairly loose terms, the precise meaning of which, together with
the true limitations of perturbation theory, will become clear in the
next section.

\begin{figure}[t]
\vspace*{-1cm}
\hspace{0cm}
\epsfysize=16cm
\centerline{\epsffile{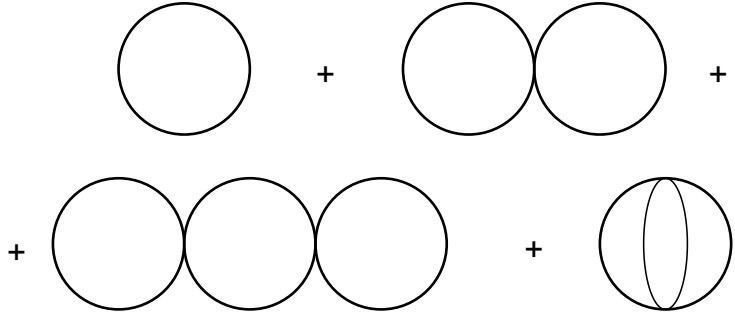}}
\vspace{-11.0cm}
\caption[a]{\protect Vacuum diagrams contributing to the 
free energy density in a scalar theory.}
\end{figure}

An attempt to compute perturbatively the statistical sum or Green's
functions at sufficiently large temperatures immediately shows the
trouble \cite{weitemp}. As the simplest example let us take the theory
(\ref{phi4}) and compute its statistical sum. In perturbation theory,
it is given by a set of vacuum graphs, see Fig. 1. Consider the
temperatures $T \gg m$. Then, with the use of Feynman rules defined
above, it is easy to see that in addition to an ordinary, 
zero-temperature expansion parameter, $\sim \lambda$, a new expansion
parameter appears, 
\be
\sim \frac{\lambda T}{m}.
\label{exppar}
\ee
A most simple way to see this is to take the $n=0$ contribution to the
statistical sum (as we will see later, these modes are crucial in
constructing effective field theories). The simple loop gives
$m^3 T$, the ``figure of eight" graph gives $\lambda T^2 m^2$, etc.
Therefore, the straightforward perturbation theory breaks down at $T >
m/\lambda$. For the theories containing, in perturbation theory,
massless bosons (such as QCD) perturbation theory does not work
for any temperature \cite{linde:pl80,gpy}.

There is a deep physical reason why this happens. At zero
temperature we apply perturbation theory for consideration of
processes where only a small number of particles participate. Thus, the
expansion parameter is $\lambda$. At high temperatures, the number of
particles, participating in collisions, may be large. Moreover, for
bosonic degrees of freedom there is a well-known Bose amplification
factor, associated with the bosonic distribution
\be
n_B(E)=\frac{1}{\exp(E/T)-1},
\ee
where $E=\sqrt{k^2 + m^2}$ is the particle energy. So, the expansion
parameter becomes $\lambda n(E)$, coinciding with
(\ref{exppar}) at small momenta.

In fact, perturbation theory breaks down in the most interesting
place, namely at the temperatures where different phase transitions
are expected. One of the ways to deal with this problem is to
rearrange the perturbative series and make a resummation of the most
divergent diagrams. Partially, this helps in some cases. For example,
for our pure scalar model the summation of the bubble graphs is
equivalent to the introduction of the temperature-dependent scalar
mass, which comes from one-loop corrections \cite{doljack} (see fig. 2):
\be
m_{eff}^2(T) = m^2 + \frac{\lambda}{4} T^2,
\ee
which is to be used in propagators. For positive $m^2$ this procedure
saves the situation and allows the perturbative computation of all
properties of the equilibrium plasma in this theory. An interesting
thing is that at high temperatures the expansion parameter of the
resummed perturbation theory  is $\frac{\lambda T}{m_{eff}} \sim
\sqrt{\lambda}$ rather than $\lambda$. For negative values of
$m^2$, corresponding to the spontaneous symmetry breaking at zero
temperatures, even resummed perturbation theory breaks down near the
point $m_{eff}^2(T)\simeq 0$, failing do describe the details of the
symmetry-restoring phase transition which occurs there.

\begin{figure}[t]
\vspace*{-0.8cm}
\hspace{2cm}
\epsfysize=16cm
\centerline{\epsffile{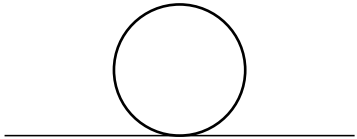}}
\vspace{-13.5cm}
\caption[a]{\protect One-loop correction to the scalar mass.}
\end{figure}

Take now realistic theories. The resummed perturbation theory does
not solve the problem, say, in QCD. The static gluon ($n=0$ component
of the gauge field) mass remains zero in the one-loop approximation, so
that infrared divergence inherent in (\ref{exppar}) is not cut.
A typical energy scale $E$, at which this happens is \footnote{The
origin of this scale will become more clear in the next
section.} $E \sim \alpha_s T$ \cite{linde:pl80,gpy} and is much
smaller than the temperature itself because, according to our
assumption, the coupling constants are small at the temperature under
consideration. The same problem appears in the description of the
phase transitions in gauge theories with scalars (such as EW
theory and GUTs), where the gauge fields at high temperature are
massless in perturbation theory.

To summarize this discussion, even the resummed perturbation theory
breaks down at some infrared energy scale $E_{infrared} \sim \alpha
T$, where $\alpha$ is a typical coupling constant of the theory under
consideration. Thus, a number of properties in high-temperature
plasma and different phenomena such as phase transitions cannot be
described by perturbation theory.

If perturbation theory breaks down a natural inclination would be
the use of direct numerical non-perturbative methods, such as lattice
Monte Carlo simulations. This approach does not work, however, for
theories containing chiral fermions, since we do not know how to put
these on the lattice. Thus, theories such as the EW
theory or grand unified models cannot be studied on the lattice with
their complete particle content. This problem does not appear in pure
bosonic models or in theories containing vector-like fermions, such as
QCD. These models can be simulated on the lattice, but 
computations are often very demanding. Quite ironically, the computations
are {\em more} time consuming for {\em weaker} coupling constants.
This can be seen as follows. At high temperatures, the average
distance between particles is of the order of $T^{-1}$, and it is clear
that the lattice spacing $a$ must be much smaller than this distance,
$a \ll T^{-1}$. At the same time, the lattice size $N a$, where $N$
is the number of lattice sites in the spatial direction, must be much
larger than the infrared scale, described above, i.e. $ Na \gg (\alpha
T)^{-1}$. Therefore, the lattice size is required to be rather large,
$N\gg \frac{1}{\alpha}$, the larger the smaller the coupling constant
is.

The next section describes the formalism which allows one to deal
with these problems, at least when the effective coupling constants
are small at the temperature under consideration. Other limitations
of the effective field theory approach will be considered later.

\section{Effective field theory approach}
The main idea of the effective theory approach to high-temperature
field theory is the factorization of weakly coupled high-momentum modes,
with energy $E \gg \alpha T$, and of strongly coupled infrared modes with
energy $E < \alpha T$, and the construction of an effective theory for
infrared modes only. The construction of the effective field theory
is perturbative, while its analysis  may be non-perturbative.
Different methods can be applied: $\epsilon$ expansion, exact
renormalization group, gap equations, Monte Carlo simulations, etc.
Thus, a combination of perturbative and non-perturbative methods is to be
used to solve the problem. The idea of this construction, known as
dimensional reduction, goes back to the papers by Ginsparg
\cite{Gins80}, and by Appelquist and Pisarski \cite{App81}. It was
developed in refs. \cite{K5,K3,K2,K1}, in application to the phase
transitions and applied to hot QCD in refs.
\cite{Bra5,Bra4,Bra1,Bra2,Bra3}. Different aspects of dimensional
reduction were studied in refs.
\cite{Jack81,Lands89,Jako94}.

The Euclidean formulation of the finite temperature field theory,
described in section 2, provides a natural recipe for the
construction of effective field theory. We learned there that the
finite-temperature equilibrium field theory is equivalent to the
Euclidean field theory  defined on a finite time interval. Let us
expand the bosonic and fermionic fields into Fourier sums,
\be
\phi(x,\tau)=\sum_{n=-\infty}^{\infty}\phi_n(x)\exp(i
\omega^b_n \tau),
\ee
\be
\psi(x,\tau)=\sum_{n=-\infty}^{\infty}\psi_n(x)\exp(i
\omega^f_n \tau),
\ee
where $\omega^b_n=2 n \pi T$ and $\omega^f_n=(2n+1) \pi T $. After
inserting these expressions into the action, the integration over time
can be performed explicitly. As a result, we get a 
{\em three-dimensional} action, containing an infinite number of fields,
corresponding to different Matsubara frequencies.
Symbolically,
\be
\int d^4x L \rightarrow \sum \int d^3 x L ^{3d}.
\ee
Therefore, a 4d finite-temperature field theory is equivalent to a 3d
theory with an infinite number of fields, and 3d boson and fermion
masses are just the frequencies $\omega^b$ and $\omega^f$. One can easily
recognize here a perfect analogy to Kaluza--Klein theories with
compact higher-dimensional space coordinates.

Now comes a crucial step. The 3d ``superheavy" modes (fields with
masses $\sim \pi T$) interact weakly with each other and with light
modes (bosonic fields corresponding to the zero Matsubara frequency).
Therefore, they can be ``integrated out" with the use of perturbation
theory, so that the 3d effective action, containing zero modes of
bosonic fields, can be constructed. Formally,
\be
\exp(-S_{eff}) = \int D\Psi D\phi_n \exp(- S_E),
\ee
where the product is taken over non-zero frequencies. The effective action
contains the bosonic fields only and can be written in the form
\be
S_{eff}= c V T^3+\int d^3x \left[L_{b}(T) + \sum_{n=0}^{\infty}
\frac{O_n}{T^n}\right],
\ee
where $L_{b}(T)$ is a super-renormalizable 3d effective bosonic
Lagrangian with temperature-dependent constants, containing a scalar
self-interaction up to the fourth power, $O_n$ are operators of
dimensionality $n+3$, suppressed by powers of temperature at $n\geq 1$,
$c$ is a perturbatively computable number (contribution of $n\neq 0$
modes to the free energy density). The existence of the effective action is closely related to the decoupling theorem \cite{decoupl,decouplms}.

A way to compute the effective action of the theory is by a matching
procedure, which has a lot in common with the corresponding method
for heavy quarks, described at this school by M. Neubert. Write the
most general 3d effective action, containing the light bosonic fields
only, and fix its parameters (coupling constants and counterterms) by
requiring that 3d Green's functions at small spatial momenta $k\ll
T$, computed with an effective Lagrangian, coincide with the initial 4d
static Green's functions up to some accuracy,
\be
G^{3d}(k_1,...k_n) = G^{4d}_{\omega=0}(k_1,...k_n)(1+O(g^m)).
\ee
Depending on the level of required accuracy, different numbers of
operators $O_n$ must be included in the effective theory. For a
generic gauge theory with $\lambda \sim g^2, f \sim g$, where
$\lambda$ ($f$)  is a typical scalar self-coupling (fermion Yukawa
coupling), $m=4$ for a super-renormalizable part of the effective
theory, i.e. when all operators $O_n$ are dropped.

The following, evident consistency check must be applied to the
constructed effective field theory. The typical energy scales (masses
of excitations in 3d theory $m_{eff}$) must be small compared with
the energy scale $\pi T$ that we have integrated out:
\be
\left(\frac{m_{eff}}{\pi T}\right)^2 \ll 1.
\ee
In fact, the 3d approximation generalizes the so-called 
high-temperature expansion often applied to the construction of the
effective potential of the scalar field at high temperatures.

After the effective field theory is constructed, the statistical sum
is expressed through the functional integral over 3d bosonic fields
only:
\be
Z=\int D\phi(x)\exp(-S_{eff}).
\label{stsum}
\ee
In some cases this integral can be computed with the help of the
perturbation theory, but in general its evaluation requires different
non-perturbative methods.

The construction, described above, looks quite formal. Nevertheless,
it has a nice physical interpretation associated with the classical
statistics of the field theory. Indeed, consider the bosonic fields
appearing in $L_b(T)$ as generalized coordinates for some classical
field theory with the Hamiltonian
\be
H= \int d^3x ~\left[\frac{1}{2}\sum P_i(x)^2 + T L_b(T) \right],
\label{class}
\ee
where $P_i(x)$ is a set of generalized momenta. Now, the partition
function for this classical system is given by the functional
integral
\be
Z= \int DP(x)D\phi(x) \exp\left(-\frac{H}{T}\right),
\ee
which coincides with eq. (\ref{stsum}) after the integration over
momenta. Thus, it is often said that the high-temperature limit of
the quantum field theory is given by the classical statistics
\cite{pol}.

We will demonstrate how this general procedure works on specific
examples in the next section. As we will see, in several cases
further simplification of the effective theory is possible.

\section{Examples}
\subsection{Pure scalar field theory}
Let us first consider a simplest scalar theory with the 4d Lagrangian
defined by eq. (\ref{phi4}). According to our rules, we must write down the
most general 3d Lagrangian, consistent with the symmetries
of the theory. It looks like:
\be
L_{eff} = \frac{1}{2}(\partial_{i}\phi_3)^2 + \frac{1}{2}m_3^2
\phi_3^2 + \frac{1}{4}\lambda_3 \phi_3^4 +\Delta L,
\label{scalareff}
\ee
where $\Delta L$ represents the contribution of higher-order
operators. In 3d, the dimensionality of the different coupling
constants and scalar field are: $\phi_3$: GeV$^{\frac{1}{2}}$;
$m_3^2$ : GeV$^2$; $\lambda_3$: GeV. The mapping procedure on the
tree level (sometimes called naive dimensional reduction) immediately
gives the relation of 3d field to 4d field: $\phi_3 = \phi/\sqrt{T}$,
and other 3d parameters are given by $m_3^2 = m^2$ and $\lambda_3 =
\lambda T$.

The structure of one-loop corrections to these relations can be
easily found on general grounds. An important fact is that the 3d
theory is super-renorma-lizable and contains a finite number of
divergent diagrams only. The ultraviolet renormalization of the
coupling $\lambda_3$ is absent in any order of perturbation theory,
while the mass term contains linear and logarithmic divergences on
the one- and two-loop levels respectively. At the same time, the
4d self-coupling $\lambda$ and mass $m$ are scale-dependent. 
Thus, on the one-loop level the relation of the 3d parameters
to the 4d ones must have the form:
\[
m_3^2 = m^2(\mu)\left[1+\beta_m \log\frac{\mu_T}{\mu}\right] + A\lambda T^2,
\]
\be
\lambda_3 = \lambda(\mu) T\left[1+\beta_\lambda \log\frac{\mu_T}{\mu}\right],
\label{phired}
\ee
where $\beta_m$ and $\beta_\lambda$ are the $\beta$-functions
corresponding to the running mass and self-coupling. The parameters
$A$ and $\mu_T$ cannot be fixed by the requirement of renormalization
group invariance and are to be found by explicit computation of
diagrams in Figs. 2 and 3. In the modified minimal subtraction
scheme ${\overline{{\rm MS}}}$ they are:
\be
\mu_T=4 \pi T e^{-\gamma}\approx 7 T,~~~ A=\frac{1}{4}.
\label{muT}
\ee
In eqs. (\ref{phired}) the parameter $\mu$ is arbitrary, and taking it to
be $\mu=\mu_T$ minimizes the corrections. This allows us to rewrite
these equations in a simpler form, $\lambda_3=\lambda(\mu_T)T$,
$m_3^2= \frac{1}{4}\lambda(\mu_T)T^2+m^2(\mu_T)$. The appearance of
the running mass and 4d coupling constant at scale $\sim T$ clarifies
our requirement concerning the amplitude of the coupling constant,
needed for the perturbative construction of the effective field theory.
The higher-order corrections to relations (\ref{phired}) can also be
found.

\begin{figure}[t]
\vspace*{-1cm}
\hspace{0cm}
\epsfysize=16cm
\centerline{\epsffile{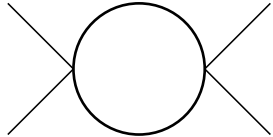}}
\vspace{-13.0cm}
\caption[a]{\protect One loop correction to the coupling constant.}
\end{figure}

The super-renormalizable theory gives the accuracy in Green's function
$\frac{\Delta G}{G} \sim O(\lambda^2)$ provided we have spatial momenta $k
\ll T$ and a 3d mass $|m_3^2| \ll T^2$. A further increase in the
accuracy may be achieved by adding higher-order operators. The one
with dimensionality three appears on the one-loop level and is equal to $h
\phi_3^6$, with
\[
h=\frac{9 \zeta(3)}{256 \pi^4}\lambda^3.
\]

Just on dimensional grounds, the perturbative expansion parameter in
the effective 3d theory is $\lambda_3/m_3$. Thus, if the spontaneous
symmetry breaking is absent at zero temperature ($m^2 > 0$), then
the 3d theory is weakly coupled in the whole range of temperatures (of
course, below the Landau pole). Thus, all equilibrium properties of a
high-temperature state can be computed by first constructing the
effective field theory and then by perturbative computations in the
effective theory. In the opposite case, when $m^2$ is negative, the
symmetry $\phi \rightarrow -\phi$ is broken at zero temperature, and
the scalar field acquires a non-zero vacuum expectation value:
\be
v^2 = -\frac{m^2}{\lambda}.
\ee
As seen from our effective Lagrangian (\ref{scalareff}), the 3d mass
changes its sign at
\be
T_c^2 = - \frac{4 m^2}{\lambda},~~~ T_c = 2 v,
\ee
which is nothing but an estimate of the critical temperature of the
phase transition with restoration of the $\phi \rightarrow -\phi$
symmetry. The mass squared of the particle excitation in the tree
approximation is given by $m_3^2$ at $T > T_c$ and by  $-2 m_3^2$ at
$T<T_c$. The perturbative analysis of the {\em effective theory} near
$T=T_c$ breaks down, since the expansion parameter $\lambda_3/|m_3|$
diverges near this point. Non-perturbative methods (such as
$\epsilon$-expansion or lattice simulations) are needed in order to
clarify the nature of the system here.

Let us consider the advantages we gain by the construction of the
effective field theory. The straightforward perturbative analysis of
the original theory is valid for $\lambda \ll 1$ and $T^2 \ll
\frac{|m^2|}{\lambda}$. The {\em construction} of the effective field
theory requires only $\lambda \ll 1$ and $|m^2| \ll (\pi T)^2$. In
particular, it is applicable near the temperature of the phase
transition $T_c$. The original 4d theory at $T^2 >
\frac{m^2}{\lambda}$ contains at least two important energy scales:
an ultraviolet one $\sim T$ and the infrared one $\sim m_3$; the
effective theory contains an infrared scale only. Beyond the phase
transition, the theory is perturbatively solvable, and the effective
field theory approach provides a convenient recipe for resumming the
perturbation theory.

\subsection{QCD}
The QCD Lagrangian for $n_f$ quark flavours is given by
\be
L={1\over4} F_{\mu\nu}F_{\mu\nu} + \sum
\bar{\psi}_i(\gamma_\nu D_\nu -m_i) \psi_i.
\label{qcdlagr}
\ee
For simplicity, let us consider the limit $T \gg m_i$.
The 3d effective Lagrangian contains gauge fields $A_i^a$ and a
scalar octet $A_0^a$, $a=1,...,8$, where $A_0$ is a temporal
component of the 4d gauge field. The most general
super-renormalizable 3d Lagrangian, containing these fields, is
\be
L_3={1\over4} F_{ij}F_{ij} +\frac{1}{2}
(D_iA_0^a)^2+\frac{1}{2}m_D^2A_0^aA_0^a+
\frac{1}{4}\lambda_A(A_0^aA_0^a)^2.
\label{qcdeff}
\ee
Here $D_i$ is a covariant derivative in adjoint representation. Of
course, on the tree level $m_A^2=0$ and $\lambda_A=0$ because of the
structure of the 4d Lagrangian. The one-loop relations in the
${\overline{{\rm MS}}}$ scheme are \cite{Lands89,Huang:1995,K2}:
\ba
g_3^2 & = & g^2(\mu)T\biggl[1+\frac{g^2}{16\pi^2}\biggl(
11 L_b- \frac{2n_f}{3} L_f+1\biggr)\biggr],\\
m_D^2 & = & g^2(\mu) T^2 \left(1 + \frac{n_f}{6}\right),\\
\lambda_A&=& = \frac{3 g^4(\mu) T}{8 \pi^2}\left(1 - \frac{n_f}{9}\right)
\ea
where
\be
L_b= 2\log\frac{\mu}{\mu_T},~~~L_f= 2\log\frac{4 \mu}{\mu_T}.
\ee
The logarithmic corrections to $m_D^2$ and $\lambda_A$ appear on
two-loop level only. The non-zero value of $m_D$ (the Debye mass) ensures
the screening of chromo-electric fields in a high-temperature plasma
and the absence of confinement.

Because of asymptotic freedom, the effective theory is valid at
sufficiently high temperatures, when $\alpha_s(\mu_T)/\pi \ll 1$. In
other words, it is applicable to the quark-gluon phase of QCD only
and describes physics on the energy scales $k \ll T$, but $k$ may be
as large as $g T$.

The 3d Lagrangian (\ref{qcdeff}) contains two essential mass scales.
The largest one is the Debye screening mass $\sim g T$, and the smallest
scale is associated with the 3d gauge coupling
constant $\alpha_3= \alpha_s T$. The scale hierarchy $\alpha_3 \ll m_D$, appearing when the
effective field theory approach is valid, suggests that a further
simplification of the effective theory is possible. Namely, the
``heavy" scale $\sim g T$ (we used the term ``superheavy" for the
scale $\sim T$) can be integrated out. The construction of the
``second" level of effective field theory, containing the
chromo-magnetic gauge fields only, goes precisely along the lines
described above. The result is a pure SU(3) Yang--Mills theory,
describing the interaction of soft modes with momenta $k \ll gT$, but
$k$ may be as large as $g^2 T$. The Lagrangian is:
\be
L_{eff} ={1\over4} F_{ij}F_{ij}.
\ee
In the one-loop approximation the new gauge coupling is \cite{K4}:
\be
\bar{g}_3^2 = g_3^2\left(1-\frac{g_3^2}{16 \pi m_D}\right),
\ee
and the accuracy of an effective description by a pure gauge theory at
momenta $k \ll gT$ is $\frac{\Delta G}{G} \sim O(g^3)$ \cite{K2}.

The 3d pure gauge theory which appeared as a final stage of
dimensional reduction contains just one scale, $\alpha_3$. It is
strongly coupled at small energies and is believed to be confining.
No perturbative methods are available for its study. Thus, 
high-temperature QCD, in spite of asymptotic freedom, contains a piece
of non-perturbative physics described by a pure Yang--Mills theory in 3d. A
simple power counting allows one to find easily the limits of
perturbation theory for the study of high-temperature QCD. Let us take,
for example, free energy. From dimensional grounds, the
contribution from the non-perturbative pure 3d sector is of the order
of $\alpha_3^3 T$. Thus, the $O(g^6)T^4$ correction to the free energy
cannot be computed perturbatively. Recently, the perturbative
computations were pushed to the very end: the $O(g^4)$ corrections to
the free energy were computed in refs. \cite{Arnold:1994,Arnold:1995} and
$O(g^5)$ in \cite{Zhai:1995,Bra2,Bra3}. In order to find the $O(g^6)$
contribution, a non-perturbative method, such as lattice Monte Carlo
simulations, must be applied to solve the pure 3d model. In addition,
a number of 4-loop computations should be done.

\subsection{Electroweak theory}
Our experience with the pure scalar theory and QCD allows us an easy 
guess of the effective 3d action for soft strongly interacting bosonic
modes with  $k \ll gT$, describing the high-temperature EW
theory. This is just the 3d SU(2)$\times$ U(1) gauge theory with the
doublet of scalar fields with the Lagrangian
\be
L  =
\frac{1}{4}G^a_{ij}G^a_{ij}+ \frac{1}{4}F_{ij}F_{ij}+
(D_i\Phi)^{\dagger}(D_i\Phi)+
\bar{m}_3^2\Phi^{\dagger}\Phi+\bar{\lambda}_3
(\Phi^{\dagger}\Phi)^2 ,
\label{univers}
\ee
where $G_{ij}^a$ and $F_{ij}$ are the SU(2) and U(1) field strengths,
respectively, $\Phi$ is a scalar doublet, and $D_i$ is a standard
covariant derivative in the fundamental representation. The four
parameters of the 3d theory (scalar mass $\bar{m}_3^2$, scalar
self-coupling constant $\bar{\lambda}_3$, and two gauge couplings
$\bar{g}_3$ and $\bar{g}'_3$) are some functions of the initial
parameters and temperature. They were computed in the one- and partially
in the two-loop approximation in refs. \cite{K5,K2}; we present here just
the tree relations for the coupling constants:
\be
g_3^2=g^2 T,~~~g_3'^2=g'^2 T,~~~ \lambda_3=\lambda T,
\ee
and the one-loop relation for the scalar mass:
\be
m_3^2(\mu) =  -\frac{1}{2} m_H^2
+T^2\biggl(\frac{1}{2}\lambda+\frac{3}{16}g^2+\frac{1}{16}g'^2+
\frac{1}{4}g_Y^2\biggr).
\label{m3}
\ee
Here $m_H$ is the zero-temperature Higgs mass, $g_Y$ is the Yukawa
coupling constant corresponding to the $t$-quark.

As usual, the effective action does not contain fermions since
their 3d masses are ``superheavy". It does not contain zero
components of the gauge fields -- triplet and singlet of SU(2) --
because these are ``heavy" according to our classification of scales.

The most interesting area of application of the effective action
(\ref{univers}) is the region of temperatures where the electroweak
phase transition is expected to occur. As in the case of the pure
scalar theory, a rough estimate of the critical temperature follows
from the requirement that the 3d mass of the scalar field is close to
zero, $\bar{m}_3^2 =0$. In the vicinity of this point the effective
Lagrangian (\ref{univers}) has a much wider area of application than
the minimal Standard Model (MSM). In fact, it plays the role of the
universal theory which describes the phase transition in a number of
extensions of the Standard Model at least in a part of their
parameter space. The set of models includes the minimal
supersymmetric Standard Model (MSSM) and some extended versions of
it, an electroweak theory with two scalar doublets, etc. One may
wonder where are the other scalars, typical for the extensions of the
Standard Model. The answer is that all extra scalar degrees of
freedom are naturally ``heavy" (mass $\sim g T$) near the phase
transition temperature and can be integrated out.

Indeed, let us take as an example the two-Higgs doublet model. The
integration over the ``superheavy" modes gives a 3d 
SU(2)$\times$U(1) theory with an extra Higgs doublet in addition to the theory
considered above. Construct now the one-loop scalar mass matrix for
the doublets and find the temperatures at which one of its
eigenvalues is zero. Take the higher temperature; this is the
temperature near which the phase transition takes place. Determine
the mass of the other scalar at this temperature. Generally, it is of
the order of $g T$, and therefore, it is heavy. Integrate this heavy
scalar out -- the result is eq. (\ref{univers}). In the case when
both scalars are light near the critical temperature, a more
complicated model, containing two scalar doublets, should be studied.
It is clear, however, that this case requires fine tuning.

The same strategy is applicable to the MSSM. If there is no breaking
of colour and charge at high temperature (breaking is possible, in
principle, since the theory contains squarks), then all degrees of
freedom, excluding those belonging to the two-Higgs doublet model,
can be integrated out. We then return back to the case considered
previously. The conclusion in this case is similar to the previous
one, namely that the phase transition in the MSSM can be described by
a 3d SU(2)$\times$U(1) gauge--Higgs model, at least in a considerable
part of the parameter space. The explicit relations were worked out
in refs. \cite{ClKa,Losada,Laine96}. What changes from going from one
theory to another is the explicit perturbative relations between
initial 4d parameters and parameters of the effective theory; if two
different 4d theories have the same 3d couplings, the electroweak
phase transition occurs in them in a similar way. The effective field
theory approach again demonstrates its strength: instead of studying
many models with different particle content, it is sufficient to study
just one 3d effective theory by non-perturbative means; the result of
this study may be used for many 4d models after {\em perturbative}
computations of 4d $\rightarrow$ 3d mapping.

\section{Electroweak phase transition}
One of the most interesting areas of application of 
finite-temperature field theory are phase transitions. Our interest in this
section will be an EW theory. The strength of the
electroweak phase transition is important for a number of
cosmological applications. For example, all mechanisms of
electroweak baryogenesis require that the phase transition should be
strong enough, i.e. \cite{s:m^14,s:sm87}
\be
v(T_c)/T_c > 1,
\label{ewconst}
\ee
where $v(T_c)$ is the vacuum expectation value of the Higgs field
at the critical temperature.

As we have learned in the previous section, to study the electroweak
phase transition it is sufficient to study an SU(2)$\times$U(1)
gauge--Higgs theory in 3d. Let us simplify  even further and
omit the U(1) factor. Numerically the U(1) coupling constant $g'$ is
smaller than the SU(2) one $g$; thus the corrections are expected to
be small. In this case, the Lagrangian is
\be
L  =
\frac{1}{4}G^a_{ij}G^a_{ij}+
(D_i\Phi)^{\dagger}(D_i\Phi)+
\bar{m}_3^2\Phi^{\dagger}\Phi+\bar{\lambda}_3
(\Phi^{\dagger}\Phi)^2 .
\label{lagrhiggs}
\ee
This 3d theory is defined by one dimensionful parameter $g_3^2 \sim
g^2 T$ and two dimensionless ratios
\be
\qquad x\equiv {\bar{\lambda}_3\over \bar{g}_3^2},\qquad
y\equiv \frac{\bar{m}_3^2}{\bar{g}_3^4}.
\label{3dvariables}
\ee
The dimensionful coupling constant can be chosen to fix the energy
scale. Therefore, the phase state of this theory is completely
defined by the two numbers $x$ and $y$. For the MSM the dependence of
the parameter $x$ on the mass of the Higgs boson near the critical
temperature (near $y=0$) is shown in Fig. 4.

\begin{figure}[t]
\vspace*{1.2cm}
\hspace{0cm}
\epsfysize=11cm
\centerline{\epsffile{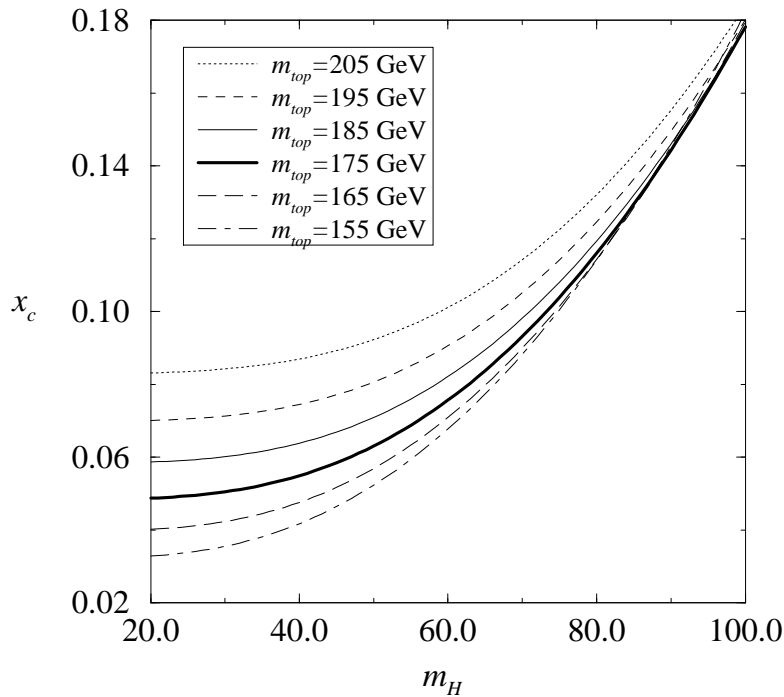}}
\vspace*{-3cm}
\caption[a]{\protect
The  value $x=\lambda_3/g_3^2$ as a function of the
physical Higgs
mass $m_H$ and the top quark mass $m_{\rm top}$ near the critical temperature defined from taking $m_3^2=0$. In general,
$x$ depends on the Higgs mass, the top mass, and logarithmically
on the temperature. From ref. \cite{K1}.}
\label{xmhdep}
\end{figure}

At first glance there are two phases in this theory. When the Higgs
mass is positive, $y>0$, one would say that this theory is analogous
to QCD with scalar quarks. Thus, we are in the ``confinement" phase
(other names of this phase are ``symmetric" or ``restored" phase) and the
particle spectrum consists of the bound states of the scalar quarks,
such as
\[
\pi = \Phi^{\dagger}\Phi, \quad W_j^{0}=i(\Phi^{\dagger}D_j\Phi-(D_j\Phi)^{\dagger}\Phi),
\]
\be
W_j^{+}=(W_j^{-})^*=(\Phi^{\dagger}D_j{\tilde\Phi}-
(D_j\Phi)^{\dagger}{\tilde\Phi}).
\label{operators}
\ee
Here $\tilde{\Phi}=i\tau_2\Phi^*$.

On the contrary, if the scalar mass is negative, one would say that
the SU(2) gauge symmetry is broken, and the initially massless gauge
bosons acquire non-zero masses. This phase is usually called ``broken"
or Higgs. In the usual folklore, the particle spectrum consists of
fundamental massive gauge bosons and the Higgs particle.

When the temperature changes, the sign of the effective mass term
changes. If the consideration presented above were true, one would
expect to have a first-order or second-order phase transition between
the two phases. In fact, the gauge symmetry is never ``broken" -- all
physical observables by construction are gauge-invariant. Moreover,
there is no gauge-invariant local order-parameter that can
distinguish between the ``broken" (Higgs) and ``restored"
(symmetric or confinement) phases \cite{Banks79,Fradk79}. The bound
states defined in (\ref{operators}) are complementary to the
``elementary" excitations in the Higgs phase. A simple exercise
shows that in the Higgs phase in unitary gauge the composite fields 
defined by eq. (\ref{operators}) are
proportional to the ``elementary" fields corresponding to the Higgs
particle and vector bosons. Thus, in a strict sense, there is no
gauge symmetry restoration at high temperatures\footnote{Contrary 
to the gauge symmetry case, global symmetries may be broken
or restored. In the pure scalar model, considered in the previous
section, the symmetry $\phi \rightarrow -\phi$ is global, so that it
is indeed restored at high temperatures, provided it was broken at
$T=0$.}, but there can be (but not necessarily are) phase transitions.
\begin{figure}
\vspace*{-1cm}
\hspace{1cm}
\epsfysize=15cm
\centerline{\epsffile{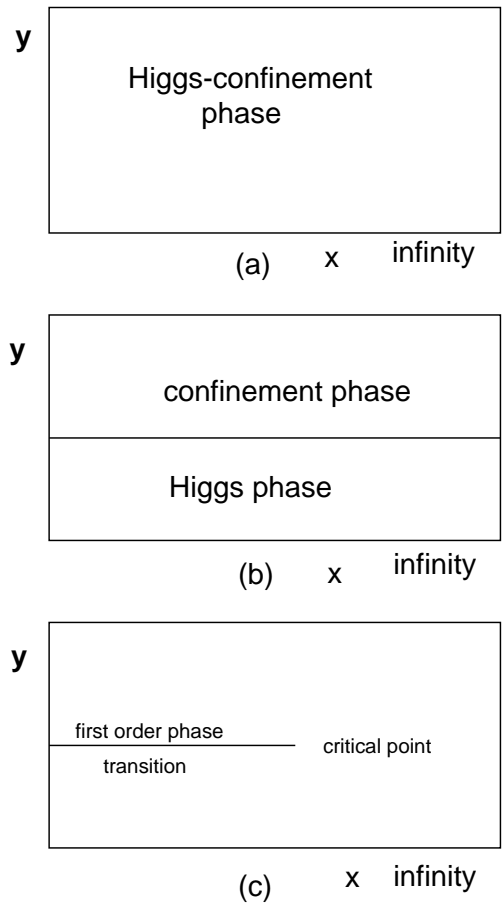}}
\vspace*{-3cm}
\caption[a]{\protect Three possible types of phase diagram for the SU(2) gauge--Higgs system.}
\end{figure}

This general consideration suggests three possible phase diagrams for
the SU(2)--Higgs model (Fig. 5). First, because the qualitative difference
between symmetric and Higgs phases is absent, we can have only one
phase (Higgs-confinement phase) everywhere at the $(x,y)$ plane (Fig.
5a). Of course, different points at this plane correspond to the
theories which are quantitatively different; nevertheless one theory
can be analytically transformed to another one. In the latter case
any high-temperature phase transition is absent. Another
possibility is that there is a {\em first-order} transition line
separating the ``symmetric" phase from the ``broken" phase. In this
case the first-order phase transition occurs at any value of the
scalar self-coupling constant (Fig. 5b).  An intermediate
possibility is when a first order phase transition line has an 
end-point somewhere on the phase plane (Fig. 5c). Then at $x< x_{crit}$
the phase transition is of the first kind, at $x=x_c$ the phase
transition is of the second kind, and at $x>x_c$ the phase transition is
absent.

Of course, some computations are necessary in order to clarify the
phase structure. A simple one-loop perturbative analysis allows Fig. 5a
to be ruled out, but cannot distinguish between Fig. 5b and
Fig. 5c.

Let us define the field-dependent vector boson mass as
\be
m_T=\frac{1}{2} g_3\phi,
\ee
and scalar masses as
\ba
&& m_1^2=m_3^2+3\lambda_3\phi^2, \\
&& m_2^2=m_3^2+\lambda_3\phi^2.\nonumber
\label{masses}
\ea
Then the 1-loop effective potential for the scalar field is
\be
V_1(\phi) = \frac{1}{2} m_3^2\phi^2
+\frac{1}{4}\lambda_3\phi^4 -{1\over12\pi} \biggl(6m_T^3 + m_1^3 +
3m_2^3 \biggr).
\label{1looppot}
\ee
This effective potential describes a first-order phase transition,
since at the critical value of $m_3^2$ the jump of the order
parameter $\phi$ is non-zero. Is the conclusion about the first-order
nature of the phase transition reliable? To answer this question, one
can estimate the value of the field $\phi$ at the {\em maximum} of
the effective potential. At sufficiently small values of $\lambda_3$
it is of the order of $ \phi_{max} \sim \frac{g_3^3}{\pi \lambda_3}$.
The dimensionless expansion parameter at this point is $\frac{g_3^2}{\pi
m_T}\sim \frac{\lambda_3}{g_3^2}=x$. Thus, the existence of the
maximum of the effective potential is reliable at small values of $x$
(small Higgs masses in zero-temperature language). Therefore, the
``symmetric" and the ``broken" phases are separated by a 
first-order phase transition line, at least at small $x$. At larger $x$ 
perturbation theory is not to be trusted, and the nature of the phase
transition can be clarified only by some kind of non-perturbative
analysis. An argument in favour of Fig. 5b follows from the
$\epsilon$-expansion \cite{Gins80,Arn94} while an indication that the
scenario of Fig. 5c may be realized comes from the study of 1-loop
gap equations \cite{Buch5}.

The 3d lattice MC simulations, done in \cite{endpoint}, have
established the absence of first-and second-order phase 
transitions at $x > 0.18$ and
singled out the phase diagram of the type shown in Fig. 5c. The value of the
end-point of the first-order phase transition line is likely to be
near $x={1 \over 8}$, i.e. there is no phase transition at 
Higgs masses greater than $80$ GeV in the minimal Standard Model. In
this case it is quite unlikely that there are any cosmological consequences
coming from the EW epoch. The 4d lattice simulations at
sufficiently small Higgs masses of a pure bosonic model were carried
out in refs. \cite{Bunk92,Bunk93,Fo94,Fo95,4d1}. Whenever the
comparison between 3d and 4d simulations is possible, they are in
agreement, indicating the correctness of the dimensional reduction
beyond perturbation theory.

The requirement of the EW baryogenesis provides an even
stronger constraint on the strength of the EW phase transition. In
fact, the constraint (\ref{ewconst}) does not hold for any Higgs mass
in the MSM (if the mass of the top quark is $175$ GeV) \cite{K1}. It is
possible to satisfy this constraint in a specific portion of the
parameter space of the MSSM \cite{cqw}: the Higgs mass is smaller than
the $Z$ mass, the lightest stop mass is smaller than the top mass,
and $\tan \beta <3$. This prediction can be tested at LEP2.

Near the critical point (the end-point of the first-order phase
transition line) the 3d gauge--Higgs system admits a further
simplification at large distances $\gg \frac{1}{g_3^2}$. At the
critical point the phase transition is of second order, thus
there is a massless scalar particle. The effective theory describing
this nearly massless state is a simple scalar theory of some field
$\chi$ with the Lagrangian
\be
L= \frac{1}{2}(\partial_i\chi)^2 + \frac{1}{2}m^2 \chi^2 +
\lambda_\chi \chi^4 +h \chi.
\ee
It is a challenge to define a mapping of the parameters of the gauge
SU(2) Higgs model to the parameters of the scalar theory, since the
perturbative methods fail in the strong coupling limit.

\section{Conclusion}
The effective field theory approach is a powerful method for studying
high-temperature equilibrium field theory. It has allowed to solve a
number of long-standing problems of high-temperature gauge theories.
The list includes the infrared problem of the thermodynamics of
Yang--Mills fields and the problem of the EW phase transition.
The method allows reliable computations of the properties of
the high-temperature equilibrium plasma of elementary particles.

Of course, the method has a number of limitations. It cannot be used
in theories where coupling constants are large at the scale of the
order of the temperature. It does not work at temperatures below the
relevant mass scales. It is not applicable to time-dependent
phenomena and cannot be used for computation of transport
coefficients such as viscosity, etc. However, it allows a description of the ground state of the system at high temperatures, therefore providing
a starting point for the study of non-equilibrium phenomena.

I am grateful to Mikko Laine and Keijo Kajantie for reading of the manuscript and helpful comments.

\section*{References}
%\bibliography{bau}
%\bibliographystyle{unsrt}

\end{document}